\documentclass[jkps,oneside,onecolumn,epsfig ]{revtex4}
\usepackage{graphicx}

\begin{document}
\title[] {Dynamical Minority Games in Futures Exchange Markets }
\author{Seong-Min \surname{Yoon}}
\affiliation{ Division of Economics, Pukyong National University,\\
Pusan 608-737, Korea}
\author{Kyungsik \surname{Kim}}
\affiliation{ Department of Physics, Pukyong National University, \\
Pusan 608-737, Korea \\}

\received{ February 2005 }

\begin{abstract}

We introduce the minority game theory for two kinds of the Korean
treasury bond (KTB) in Korean futures exchange markets. Since we
discuss numerically the standard deviation and the global
efficiency for an arbitrary strategy, our case is found to be
approximate to the majority game. Our result presented will be
compared with numerical findings for the well-known minority and
majority game models.\\
\hfill\\
$PACS$: 05.20.-y, 89.65.64, 84.35.+i\\
$Keywords$: Minority game, Standard deviation, Korean treasury
bond \\
$^{*}$Corresponding author. Tel.:+82-51-620-6354; fax:+82-51-611-6357.\\
$E-mail$ $address$: kskim@pknu.ac.kr (K.Kim).

\end{abstract}

\maketitle

\section { Introduction }

More than one decade, the minority game $[1,2]$ is a simple and
familiar model that has received considerable attention as one
interdisciplinary field between physicists and economists. There
has primarily been concentrated on calculating and simulating in
various ways for the game theories such as the evolutionary
minority game $[3-5]$, the adaptive minority game $[6,7]$, the
multi-choice minority games $[8]$, the ${\$}$-game model $[9]$,
and the grand canonical minority games. Challet et al $[10]$ have
introduced the stylized facts that the stock prices are
characterized by anomalous fluctuations and exhibited the fat
tailed distribution and long range correlation. Moreover, it
reported from previous works $[11-14]$ that the grand canonical
minority game reveal the important stylized facts of financial
market phenomenology such as return changes and volatility
clusterings. Ferreira and Marsili $[11]$ studied the dynamical
behavior of statistical quantities between different types of
traders, using the minority games, the majority game, and
 the $\$$-game.

Recently, researchers have treated mainly with Arthor's bar model,
the seller and buyer's model in financial markets, and the
passenger problem in the metro and bus, etc. They have considered
several ways of rewarding the agent's strategies and compared the
resulting behaviors of the configurations in minority game theory.
De Almeida and Menche $[15]$ have also investigated two options
rewarded in standard minority games that choose adaptive genetic
algorithms, and their result is found to come close to that of
standard minority games. Kim et al $[16]$ analyzed the minority
game for patients that is inspired by de Almeida and Menche's
problem.

Recent studies based on approaches of statistical physics have
applied the game theory to financial models. To our knowledge, it
is of fundamental importance to estimate numerically and
analytically the Korean options. In this paper, we present the
minority game theory for the transaction numbers of two kinds of
KTB in the Korean futures exchange market. For the sake of
simplicity, we limit ourselves to numerically discuss the standard
deviation and the global efficiency for particular strategies of
our model, using the minority game payoff. In Section $2$, we
discuss market payoffs and statistical quantities in the game
theory. We present some of the results obtained by numerical
simulations and concluding remarks in the final section.

\section {Minority and majority games }

First of all, we will introduce the dynamical mechanism of both
the minority game and the majority game. We assume that the agents
$N$ can decide independently whether to buy or to sell the stock
at round m. When the information and the strategy take,
respectively, the value $\mu(t)$ and $S$ at time $t$, the action
of the $i$-th agent is presented in terms of ${a_{i,s}}^{\mu(t)}
(t)$. one agent can submit an order ${a_{i,s}}^{\mu(t)} (t) =1$
(buy) or ${a_{i,s}}^{\mu(t)} (t) =-1$ (sell), and the aggregate
value, i.e., the sum of all action of agents, is given by $A(t)=
\sum_{i=0}^{N} {a_{i,s}}^{\mu(t)} (t) $. The payoffs
${U_{i,s}}^{mn} (t)$ and ${U_{i,s}}^{mj} (t)$ for minority game
and majority game models are, respectively, represented in terms
of

\begin{equation}
{U_{i,s}}^{mn} (t+1) = {U_{i,s}}^{mn}(t) - {a_{i,s}}^{\mu(t)} A(t)
\label{eq:a1}
\end{equation}
and
\begin{equation}
{U_{i,s}}^{mj} (t+1) = {U_{i,s}}^{mj}(t) + {a_{i,s}}^{\mu(t)} .
A(t). \label{eq:b2}
\end{equation}
%

\begin{figure}[]
\includegraphics[angle=90,width=7.5cm]{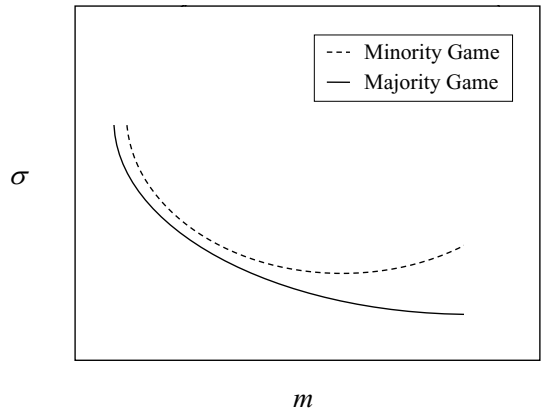}
\caption[0]{Plot of functional forms for the standard deviation
$\sigma$ as a function of the round $m$ in the minority game and
the majority game models. The standard deviation has an optimal
value in the minority game model while it takes the decreasing
value near zero as the round $m$ goes to large value in majority
game model.}
\end{figure}

From the aggregate value $A(t)$, the standard deviation of $A(t)$
is defined by

\begin{equation}
\sigma = [{\overline{\frac{1}{N} \sum_{i=1}^{N} <({A}^{2}
(t)>}}]^{1/2}, \label{eq:c3}
\end{equation}
where the bar denotes the average taken over realizations. The
statistical quantities $\sigma^{2}$ and $\sigma^{2} /N $ are,
respectively, the volatility of $A(t)$ and the global efficiency.


%
To evaluate the statistical quantity $\Phi= <A(t)A(t+1)>/<A^2
(t)>$, i.e., the ratio of autocorrelation to the volatility, two
points of view are characterized as follows: The agents become the
fundamentalists, who believe that the stock price fluctuate around
the equilibrium value for the case of $\Phi>0$, included the
minority group and the payoff of Eq. $(1)$. On the contrary, the
agents called chartists, who believe that the stock price has a
trend followed as $\Phi<0$, included the majority group and the
payoff of Eq. $(2)$. In the $\$$-game, its numerical behavior of
the statistical quantities such as the autocorrelation and the
volatility are well known to follow the minority game or the
majority game.

In order to assess the dynamical behavior for the minority game,
the majority game, and the $\$$-game, we can extend to obtain
statistical quantities such as the standard deviation, the
volatility, the auto
correlation, the self-overlap, and the
entropy, etc. We expect that these statistical quantities lead us
to more general results. Generally, Fig. $1$ shows functional
forms for the standard deviation as a function of the round $m$ in
both the minority game and the majority game.

\section { Numerical results and concluding remarks }

To estimate numerically the standard deviation and the volatility,
N agents choose one among two possible options, i.e., $1$ or $-1$
( buy and sell ) at each round m. The agents obtain the score $+1$
$(-1)$ if the return belongs to be smaller (larger) than zero. The
agent's action behaves independently without any communication or
interaction. All available information is other agent's actions,
that is, the memory of the past $m$ rounds. There are $2^m$
possible different memories for each of the $m$ rounds, and all
different strategies is well known to contain $2^{2^m}$ values.
Our model can extend to several values of strategy, but we limit
ourselves to only two strategies $S=2$ and $S=4$.

\begin{figure}[]
\includegraphics[angle=90,width=7.5cm]{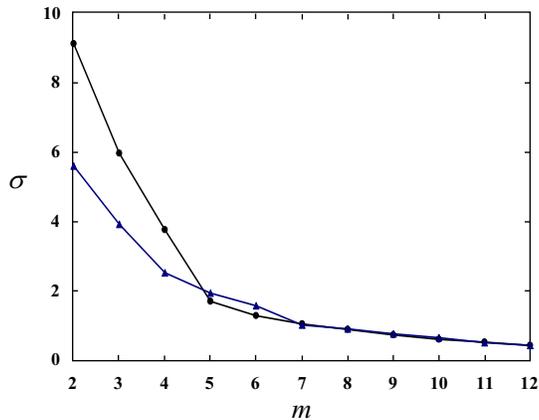}
\caption[0]{Plot of the standard deviation $\sigma$ versus m for a
strategy $S=2$ for KTB$309$ (circle) and KTB$312$ (triangle).}
\end{figure}
\begin{figure}[]
\includegraphics[angle=90,width=7.5cm]{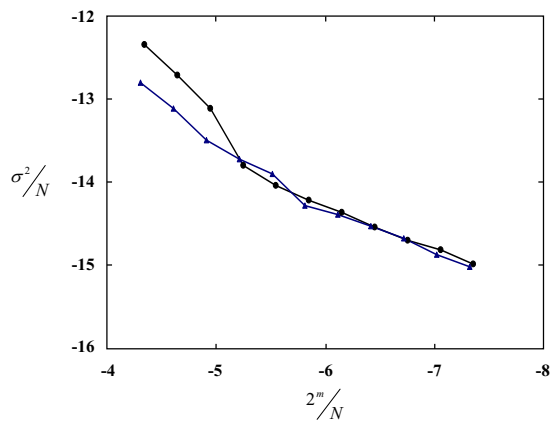}
\caption[0]{Log-log plot of the global efficiency $\sigma^{2} /N$
versus $2m/N$ for the strategy $S=2$ for KTB$309$ (circle) and
KTB$312$ (triangle).}
\end{figure}

From now, we introduce tick data of KTB$309$ and KTB$312$
transacted in Korean futures exchange market. We here consider two
different delivery dates: September KTB$309$ and December
KTB$312$. The tick data for KTB$309$ were taken from April $2003$
to September $2003$, while we used the tick data of KTB$203$
transacted for six months from July $2003$. The Korean futures
exchange market opens for 6 hours per one day except for weekends,
and the total tick data of one-minutely transactions are,
respectively, about $N=1.8\times 10^4 $ (KTB$309$) and
$N=1.9\times 10^4$ (KTB$312$). For two kinds of KTB in Korean
futures exchange market, we found results of standard deviation
$\sigma$ versus $m$ for the strategy $S=2$ in Fig. $2$, after we
estimate numerically the volatility from Eq. $(3)$. Our result is
expected to behavior in a way which is similar to the minority
game, but the dynamical behaviors for standard deviation and the
global efficiency is found to be similar to the patterns of the
majority game. The global efficiency is found to take the
decreasing value near zero as the round $m$ goes to large value,
as shown in Fig. $3$. For $S=4$, the standard deviation and the
global efficiency are also found to take similar values of Figs. 3
and 4.

In conclusions, our case is found to be approximate by the
majority game model for two kinds of the KTB$309$ and KTB$312$. It
is really found that the dynamical behavior of the standard
deviation and the global efficiency for our model is similar to
those for the majority game while the El Farol bar model $[1]$ and
the patient model $[16]$ belong to a class of the minority game.
In future, it is expected that the detail description of the
minority game theory will be used to study the extension of
financial analysis in other financial markets.

\begin{acknowledgements}
This work was supported by Korea Research Foundation
Grant(KRF-2004-002-B00026).
\end{acknowledgements}

\end{document}